\documentclass[preprint,showpacs,preprintnumbers,prb]{revtex4} 
\usepackage{graphicx}
\usepackage{amsmath}
\usepackage{times}
\usepackage{epsfig}
\usepackage{color}
\usepackage{ulem}   
\begin{document}
\def\salto{\vskip 1cm} \def\lag{\langle} \def\rag{\rangle}
\newcommand{\redit}[1]{\textcolor{red}{#1}}
\newcommand{\blueit}[1]{\textcolor{blue}{#1}}
\newcommand{\magit}[1]{\textcolor{magenta}{#1}}
\newcommand{\mylabel  }[1]{\label{#1} 
}
\title{Generalizing the self-healing diffusion Monte Carlo approach to finite temperature: a path for the optimization of
 low-energy many-body bases}
 
\author{Fernando A. Reboredo}    
\author{Jeongnim Kim} 
\affiliation {Materials Science and Technology Division, Oak Ridge
National Laboratory, Oak Ridge, TN 37831, USA}
  
\begin{abstract}
A statistical method is derived for the calculation  of thermodynamic properties of many-body systems at low temperatures.
This  method is based on the self-healing diffusion Monte Carlo method for complex functions [F. A. Reboredo J. Chem. Phys. {\bf 136}, 204101 (2012)] 
and some ideas of the correlation function Monte Carlo approach [D. M. Ceperley and B. Bernu, J. Chem. Phys. {\bf
    89}, 6316 (1988)]. 
In order to allow the 
evolution in imaginary time to describe the density matrix, we remove the fixed-node restriction  
using complex antisymmetric guiding wave functions. 
In the process we  obtain  a   parallel  algorithm that 
optimizes a small subspace of 
the many-body Hilbert space to have maximum overlap with the subspace spanned by 
the lowest-energy eigenstates of a many-body Hamiltonian. We show in a model system that the partition function is progressively  maximized
 within this subspace. We show that the subspace spanned 
by the small basis systematically converges towards the subspace spanned by the lowest energy eigenstates. 
Possible applications of this method to calculate the thermodynamic properties of many-body systems 
near the ground state are 
discussed. The resulting basis can be also used to accelerate the calculation of the ground or excited states with 
Quantum Monte Carlo. 
\end{abstract}

\date{\today}

\maketitle
\section{Introduction}

There is a significant  interest in  thermodynamical properties observed as $T \rightarrow 0$. Many physical phenomena that cover superconductivity,  magnetic and structural transitions,  chemical reactions etc. require an adequate treatment of thermal effects. These effects are crucial in  systems where there is a large number of low-energy excitations within an energy window $1/\beta=k_B T$ above the ground state.  Electronic thermal effects are expected to be larger in metals and magnets than in insulators.~\cite{QTS} In metals there is a significant number of excitations with vanishing energy. The  magnetic excitations energies frequently go to zero in the long wave limit. 
A significant fraction of spectroscopic techniques probe the electronic or magnetic excitations near the ground state. The development of first-principles techniques to obtain excitations has historically received a significant theoretical attention.~\cite{runge84,onida02,filippi09}  Monte Carlo methods used to calculate excitation energies\cite{filippi09} will be accelerated with basis that retain the physics at the relevant energies.

A first-principles finite-temperature description of many-body systems is also relevant to describe chemical reactions.\cite{mazzola12} Ionic dynamics are usually calculated within the  Born-Oppenheimer approximation. This decouples the  wave function of the ``quantum" electrons from the wave function of the ions.  Within this standard approximation, electrons are at zero temperature while the ions can move with kinetic energies that often exceed the electronic excitations. Even within the Born-Oppenheimer-ground-state approximation, the standard approach based on density functional theory (DFT) shows significant differences with diffusion Monte Carlo (DMC)~\cite{grossman97,saccani13} or Quantum Chemistry benchmarks.
 At the transition saddle points, when some chemical bonds are broken and new ones are formed, the spacing of the corresponding electronic eigenenergies is minimum, or even zero at the conical intersections.~\cite{levine07}  Electronic thermal effects are seldom included in many-body calculations.~\cite{morales10,mazzola12} In order to routinely  include thermal effects, significant improvements in the theory beyond the standard approach are required.

Most  ab-initio calculations in the literature of condensed matter electronic structure are based in the ground state quantum Monte Carlo calculations of the
homogeneous electron gas\cite{ceperley80} which made possible the first approximations of DFT.~\cite{hohenberg,kohn}
DFT has been extended to finite temperature long ego.~\cite{mermin65, kohn83,weinert92}  Fermi occupations  of Kohn-Sham eigenstates and 
the addition of an entropy term have been shown\cite{weinert92} to provide a variational density functional. However, even nowadays, the zero temperature approximation 
for the exchange-correlation potential is widely used. This approach has been long known to severely underestimate the critical Curie temperatures of magnetic systems.~\cite{statunton84,gyorffy85,staunton85}    
Including temperature for magnetic systems is possible for cases where the magnetic excitations can be treated classically~\cite{wang95,landau04} and the electrons can be assumed 
to be in the ground state for constrained configurations of the spins.~\cite{eisenbach09} But an adequate description of the electronic entropy in the subspace that preserves the spin is still lacking.~\cite{yin12} Finite temperatures benchmarks of a quality comparable to Ref.
\onlinecite{ceperley80} are  the key ingredients required to parametrize a finite temperature density functional. Without a reliable approximation,  
most work done under a DFT framework still uses a zero temperature approximation for the exchange correlation functional.

Accurate many-body calculations at high temperatures can be performed within the path integral Monte Carlo approach (PIMC).~\cite{ceperley95}  Since the cost of PIMC diverges as $T \rightarrow 0$, it has been mainly used in the hot and dense regime,~\cite{magro96,militzer00,driver12} with a temperature comparable to the interaction potential.~An alternative approach that could start from the zero temperature limit would be desirable. 

The most accurate techniques to describe a large number of electrons ($N_e>1000$) at zero temperature are based in  projection 
approaches.~\cite{ceperley80,purwanto09,booth12} One could potentially extend
 these methods to finite temperature, limiting the projector  
 $e^{-\beta \mathcal{\hat H}}$ to finite $\beta$. Thermodynamical averages can be later obtained from derivatives 
of the Helmholtz free energy
$F(\beta)=-1/\beta \ln[ tr(e^{-\beta \mathcal{\hat H}} )]$, where $\mathcal{\hat H}$ is the many-body Hamiltonian operator   and 
$tr(X)$ the trace of $X$ over the complete many-body Hilbert space. 

The standard diffusion Monte Carlo Method with importance sampling (DMC)~\cite{ceperley80,mfoulkesrmp2001,austin12} constrains
 the sign or the phase of the wave function  by imposing the nodes or the phase~\cite{ortiz93} 
 of a guiding wave function $\Psi_T({\bf R})$, where the many-body coordinate ${\bf R}= \{{\bf r}_1,{\bf r}_2,\cdots,{\bf r}_{N_e} \}$ is the set of coordinates of $N_e$ electrons. These constraints while enforcing an anti-symmetric fermionic wave function introduce a variational error. 
 The quality of the wave 
function and its nodes can be improved with several methods within a variational Monte Carlo (VMC) 
context,~\cite{mfoulkesrmp2001,ortiz95,jones97,guclu05,umrigar07,toulouse08,petruzielo12,rios06} or 
at the DMC level.~\cite{ceperley88,jones97,keystone,rollingstones,stonehenge} 
In DMC, the energy of the ground state is exact if the exact nodes or 
phase are provided.~\cite{anderson79,ortiz93}  Improving the nodes is computationally intensive. Avoiding this cost is key for finite temperature
calculations.

In standard DMC calculations, $\mathcal{\hat H}$ is replaced
by  the fixed-node Hamiltonian $\mathcal{\hat H}_{FN}$ or the fixed-phase Hamiltonian $\mathcal{\hat H}_{FP}$. 
The use of the fixed-node or fixed phase approximation can have undesired effects on the calculation of thermal effects. 
It has been found that many fermionic systems have a ground state with two nodal pockets.~\cite{bressanini12} That is, if the
ground state wave function is real the nodal surface separates the Hilbert space in only two pockets for positive and negative values
respectively. It
has been conjectured~\cite{ceperley91}  that this is a general property of fermionic ground states. 
In the fixed-node case, the excitations of $\mathcal{\hat H}_{FN}$ are forced to share the nodes  of
the ground state. To be orthogonal to the fixed-node ground state, the fixed-node excited states have to have at least an additional node.  It is known, however, that in many systems there are several fermionic excited states near the ground state with also two nodal pockets.\cite{rasch12} Accordingly, $tr(e^{-\beta {\mathcal{\hat H}}_{FN} })$ does not describe the low temperature physics. It is easy to see that the same happens in the fixed-phase case. 
Therefore, if one wishes to use a DMC-like algorithm to obtain thermodynamical properties, one must go beyond the usual
fixed-node or fixed-phase approximations. For practical reasons, a parallel approach that can handle a large number of excitations near the ground state would also be beneficial. 

In this paper, we restart the debate on how to calculate low temperature properties within a many-body ab-initio context taking into account recent
theoretical developments.~\cite{keystone,rockandroll,stonehenge}  A  method is derived that introduces temperature within an importance sampling 
procedure that shares most of the computational tools developed for projection MC approaches. 
The errors in the evolution operator $e^{-\beta \mathcal{\hat{H}}}$ resulting from the fixed-node 
restriction are eliminated by using complex 
linear combinations of eigenstates, which do not have nodes except at the electronic coincidental points 
(see Fig \ref{fg:scheme}). Instead of optimizing a single many-body 
wave-function so it better describes the ground state of the system, a basis of several wave functions is optimized   to maximize the overlap with the small subspace spanned by the lowest energy eigenstates of the many-body Hamiltonian.
We show in a model system that the overlap of the optimized subspace with the lowest energy subspace calculated
with a configuration interaction (CI) approach, increases systematically as the iterations increase and that the
partition function is maximized.  
 
The rest of the paper is organized as follows: In Section \ref{sc:theory} we describe the general formalism; some of the formulae developed in Ref. \onlinecite{stonehenge} for complex wave function is repeated here for completeness. 
 Section
\ref{sc:algorithm}  outlines the basic algorithm. 
In Section \ref{sc:results} we describe the results for a model calculation; and finally, in Section \ref{sc:summary} we
discuss the possible applications and summarize. This paper also has three appendices: 
 \ref{ap:beyondloc} describes how to go beyond the locality and local-time approximations;  \ref{ap:eigenstates} describes how to take advantage of the eigenstates when they are complex; Finally,  
\ref{ap:pairs} describes how to work with eigenstate pairs to minimize the variance of the weights of the walkers while 
keeping the wave function complex.

\section{A low-energy  expansion of the partition function}
\label{sc:theory}

This section extends the DMC approach~\cite{ceperley80} for the calculation of the partition function of a many-body system. We first provide background material required to understand the rest of the paper. We generalize the upper bound property of the energy in DMC to an upper bound property of the free energy. We next give the general outline of our approach and describe how to avoid the fixed-node approximation in DMC. Finally, we describe the details: basic formulae and numerical approach.    

\subsection{
The upper bound property of the truncated Helmholtz free energy}
Thermal effects can be obtained by calculating all excitations within a thermal energy window 
above the ground state larger than $\Delta E \sim 1/\beta = k_B T$ and then evaluating the
density matrix~\cite{reichl} as:
\begin{align}
\label  {eq:knownrho}
\hat{\rho}(\beta) = & e^{-\beta\mathcal{\hat H}} \\
= & \sum_n \left | \Psi_n \right \rangle e^{-\beta E_n} \left \langle \Psi_n \right | \nonumber ,
\end{align} 
 where$E_n< E_0+\Delta E$ is the eigenvalue with eigenvector $| \Psi_n \rangle $ of $\mathcal{\hat H}$. In general $\mathcal{\hat H}$
is given by
\begin{equation}
\label  {eq:H}
\hat{\mathcal{H}}= \sum_j^{N_e} \frac{
(\nabla_j+{\bf A}_j)^2}{2}+\hat{V}({\bf R}) 
\end{equation}
where ${\bf A}_j= {\bf A}({\bf r}_j)$ is a vector potential at point
${\bf r}_j$ with magnetic field ${\bf B}({\bf r}_j)=\nabla_j \times {\bf A}_j $, and $\hat{V}({\bf R})$ includes the electron-electron interaction, the interactions of the electrons spins with the magnetic field
and any external potential, local or non-local.

In a closed system that can exchange energy with a bath or reservoir (canonical ensemble) all thermodynamical averages 
can be obtained using the density matrix.
The trace of the density matrix $Z(\beta)=tr[\hat{\rho(\beta)}] $ is the partition function, whereas
$F(\beta)=-1/\beta\ln[Z(\beta)]$
 is  the Helmholtz free energy. 
 
 In general, $\mathcal{\hat H}$ has an infinite number of eigenvectors $| \Psi_n \rangle$  that can be ordered 
 with increasing eigenenergy $E_n$. 
If $\beta (E_n-E_0) \gg 1$, the contribution to $Z(\beta)$ of the eigenstate $| \Psi_n \rangle$ becomes negligible. Therefore, a usual approximation is to truncate the trace to a finite matrix with a finite number of eigenstates $M_S$. 

In what follows we defined $tr()$ 
as the trace of a truncated square matrix with size $M_S$. We also relate $Z(\beta)$ and $F(\beta)$ to that truncated trace. 
Since  $e^{-\beta \mathcal{\hat H}}$  is positive definite, for a given basis, $Z(\beta)$ increases 
and $F(\beta)$ decreases as $M_S$ increases. 

The trace of any linear operator is invariant for linear transformations of the form ${\hat B} e^{-\beta \mathcal{\hat H}}{\hat B}^{-1} $ 
with ${\hat B} {\hat B}^{-1} =1 $. Thus, in principle, one does not need to obtain the eigenstates 
of $e^{-\beta \mathcal{\hat H}}$ or equivalently $ \mathcal{\hat H}$ to calculate 
the free energy. Any linearly independent basis that spans the same subspace can be used to obtain $Z(\beta)$. 
Thermodynamical properties only require us to evaluate $Z(\beta)$ in a linearly independent basis $\{|\chi^{S}_m \rangle \}$. However, if statistical methods are used, then each element contributing to the truncated trace also increases the statistical error bar.
Therefore, it is computationally more efficient to use the most compact basis, with minimum $M_S$, that retains  the low-energy properties.

Any eigenstate 
$|\Psi_n \rangle$ can be written in a complete basis ${|\chi^{S}_n} \rangle$ as
\begin{align}
|\Psi_n \rangle=e^{-\hat J} \left[\sum_{m=0}^{M_S-1} \lambda_{n}^{m}  |\chi^{S}_m \rangle + \sum_{m=M_S}^{N_B\rightarrow \infty}  \lambda_{n}^{m}  |\chi^{S}_m \rangle \right],  
\end{align}
where $\langle {\bf R}|e^{-\hat J}|{\bf R}\rangle=e^{-J({\bf R})} $ is a Jastrow factor that introduces adequate cusp conditions~\cite{kato1957} and are
 $\langle {\bf R} |\chi^{S}_m \rangle=\chi_{m}^{S}({\bf R})$  linear combinations of an infinite orthogonal set $\{ \langle{\bf R}|  n \rangle \}$ [e.g. Slater determinants, or
 Pfaffians~\cite{bajdich08}, or symmetry constrained functions (SCF), etc]. In practice we restrict the Hilbert space to a finite number $N_B$. We denote the subspace spanned by
$N_B \gg M_S$ functions $|m \rangle$ as the {\it big subspace}. The big subspace has to be large enough to describe the low temperature physics of  the complete Hilbert space, which is in general infinite. We define the {\it small subspace} as the subspace spanned by the first $M_S$ basis functions $ | \chi^{S}_n \rangle $.   

Within the small basis $\{|\chi^{S}_m \rangle \}$, the free energy will be minimum if all $|\Psi_{n < M_S} \rangle$ can be spanned in the small basis, namely, 
$\lambda_{n \ge M_S}^m=\lambda_{n}^{m \ge M_S} = 0$. Errors in the small basis will result in higher values of the
free energy.  
Thus {\it the free energy in the truncated basis is an upper bound to the true Helmholtz free energy.} 
Optimization of the Helmholtz free energy in the small basis is analogous to the variational principle of the ground state. Likewise, {\it the partition function in the truncated basis is a lower bound of the exact partition function}. Improved bounds may be obtained with a basis that better describes the lower energy eigenstates. $M_S$ has to be large enough to include all the relevant physics for a given temperature.

 Most of the optimization methods in the QMC literature focus, on optimizing the eigenstates. Several methods have been proposed to obtain low-energy excited states within the linear method Monte Carlo\cite{filippi09,toulouse12} (LMMC) or diffusion Monte Carlo.~\cite{ceperley88,rockandroll}
However, since the eigenstates are sometimes  difficult calculate, and we only need an average, we argue that one might save 
computational time by optimizing 
the many-body basis first as in the correlation function diffusion Monte Carlo (CFDMC)~\cite{ceperley88} method.
Optimizing the basis directly could be more practical than obtaining accurate eigenstates energies, if the number of excitations near the ground state is large (e. g. typically the case in metallic or magnetic systems).

\subsection{Guiding ideas of the finite temperature SHDMC method and definitions}

Instead of performing the usual projection for infinite imaginary time of a single trial wave-function, we run DMC for multiple guiding wave-functions (forming a linearly independent basis) for finite imaginary time, which is equivalent to finite temperature. Instead of using a single real guiding function 
with nodes, we use a set of complex antisymmetric guiding functions without nodes.  Therefore, the Hamiltonian $\mathcal{\hat{H}}$ is not altered at the nodes as in the standard importance sampling 
DMC approach\cite{ceperley80}  with 
the fixed-node approximation.\cite{anderson79}  As explained in the introduction, extending  
DMC  to finite 
temperatures requires to go beyond those fixed schemes. Complex-valued
antisymmetric wave functions, that do not have nodal pockets, can 
be constructed as a linear combination of two real wave function with different nodes (see Fig. \ref{fg:scheme}). We 
go beyond the standard fixed-phase approximation and the local-time approximation.~\cite{stonehenge} As in the
 SHDMC method for complex wave functions,~\cite{stonehenge} the 
infamous 
sign problem is avoided with complex antisymmetric guiding functions. The result is acurate as long as enough statistical information is collected.

SHDMC\cite{keystone,rockandroll,stonehenge} systematically improves a trial wave function by  maximizing 
the overlap with the  wave function propagated in imaginary time in DMC. 
 Here instead of maximizing the overlap of a single wave function we will maximize the overlap with the basis. 

Following the ideas of the SHDMC method, we use a recursive approach. In every iteration $\ell$,  importance sampling DMC~\cite{ceperley80} is performed and statistical 
data of the evolution in $\beta$ of a set of $M_S$ guiding wave functions is projected on the many-body bases $\{ | \chi^{S,\ell}_n \rangle \}$ 
and $\{ | n \rangle \}$ 
.  The statistical data is used to improve the small basis $\{ | \chi^{S,\ell+1}_n \rangle \} $ and the guiding functions for 
 the next iteration  
In what follows, we will omit the iteration index $\ell$ in the notation for clarity, when the basis is not changed.
The small basis $\{ | \chi^{S}_n \rangle \}$ is  orthonormal:
  $\langle \chi^{S}_n  | \chi^{S}_n \rangle = \delta_{n,m}$. 


For numerical efficiency, depending on the problem, we choose guiding functions related to the eigenstates of  $\mathcal{\hat{U}}=e^{\hat{J}} e^{\mathcal{\hat{H}}}e^{-\hat{J}} $. (i) 
$\{ | \chi^{U}_n \rangle \}$ is formed by the Slater expansions of the eigenstates in the small basis. 
(ii)  $\{ | \chi^{V}_n \rangle \}$ is formed by the Slater expansion of linear combinations of  eigenstates pairs.

We construct  wave functions of the form
\begin{align}
\Psi^{T}_n({\bf R}) = &e^{-J({\bf R})} \chi^{X}_n({\bf R}); 
\label  {eq:trial}
\end{align}
where the
super index $X$ refers to either $S$, $U$, or  $V$, depending on the case. To simplify the notation, we omit $X$ in
$\Psi^T_n({\bf R}) $. 
 
We assume the Jastrow factor operator $e^{-\hat{J}}$ to be diagonal in the many-body configuration space ${\bf R}$, and  positive, which implies that it must have an inverse.~\cite{fn:back_flow} The Jastrow factor is fixed in SHDMC but can 
be optimized variationaly so that the free energy of the system is minimized. 

In contrast with the CFDMC~\cite{ceperley88} and the released phase~\cite{jones97} methods, we use anti-symmetric guiding functions, which are improved recursively
with a maximum overlap criterion. Since the exponential growth of the bosonic ground state is prevented by the guiding functions, the free energy 
obtained is an upper bound.  This approach is different to the correlated linear method~\cite{filippi09} because the wave function is optimized at the DMC level
and we use anti-symmetic guiding functions. In variance with the original SHDMC approach for excited states~\cite{rockandroll}
multiple wave functions
are propagated in parallel. A serial orthogonalization step in the original SHDMC method for 
excited states~\cite{rockandroll,stonehenge}  is postponed in this new approach until DMC has been run for the all basis functions. \cite{fn:complications}

\subsection{Working with complex guiding wave functions to avoid the fixed-node approximation}

\begin{figure}
\includegraphics[width=1.00\linewidth,clip=true]{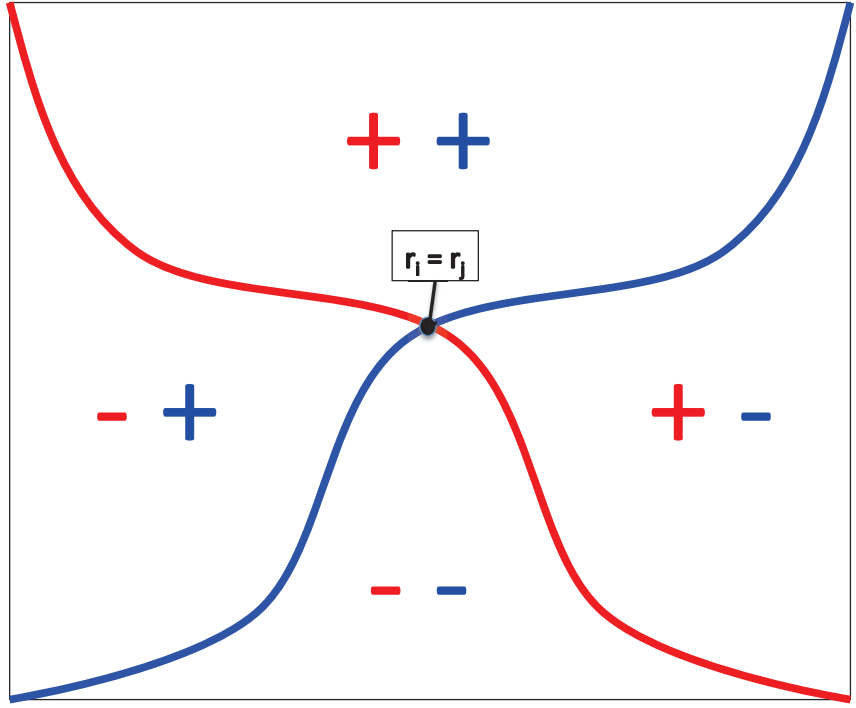}
\caption{
\label{fg:scheme} (Color online)
A schematic representation of the nodes of the real and imaginary parts of a  complex antisymmetric function. This type of 
complex functions must be zero only at the coincidental points. 
Though  real antisymmetric functions must have nodal pockets a complex linear combination must not.  
 Note also that two different antisymmetric real wave functions  approaching different eigenstates 
will have different nodes and produce different fixed-node Hamiltonians. 
}
\end{figure}


While complex guiding wave functions allow us to avoid the fixed-node approximation, they introduce additional complications~\cite{stonehenge} that are 
discussed here. Once these complications are dealt with, the sign problem is avoided as in the fixed-node. As in 
standard SHDMC the result is accurate as long as enough statistics is obtained. 

%
  
 Following Refs.~\onlinecite{ortiz93} and \onlinecite{stonehenge}, $\Psi^{T}_n({\bf
  R})=\langle {\bf R} | \Psi_n^{T} \rangle$ can be written as an explicit
product of a complex phase and an amplitude $\Psi^{T}_n({\bf
  R})=\Phi_n^{T}({\bf R}) e^{{\bf i} \phi_n({\bf R})}$. 
  
The expressions
\begin{align}
\label  {eq:defPhi}
\Phi_n^{T}({\bf R}) & =  \sqrt{\Psi_n^{T}({\bf R})\Psi^{T *}_n({\bf R})   }, \text{    and } \\
\label  {eq:phaselog}
\phi_n({\bf R}) &= \ln[\Psi_n^{T}({\bf R})/\Psi_n^{T *}({\bf R}) ]/ (2 i) +  \pi m 
\end{align}
allow the computation of all the
gradients and Laplacians in terms of those of an arbitrary complex function $\chi_n^{X}({\bf R})$
and  $e^{-J({\bf R})}$. 

   In Eq. (\ref{eq:phaselog}) $m$ is an arbitrary integer that
changes the Riemann branch of the natural logarithm $\ln$ of a complex number. $m$ only contributes
to the gradient or Laplacian at the Reimann cuts.~\cite{fn:reimann} Since the position of the Reimann cuts is an arbitrary mathematical convention,
their contribution to gradients and Laplacians is unphysical and ignored.   

The dependence in $\beta$ of
$e^{-\beta \mathcal{\hat H}} \Psi_n^{T}({\bf R})$ is given by
\begin{align}
\Psi_{n}^{T}({\bf R},\beta) = & \; \label  {eq:U}
e^{-\beta \hat{\mathcal{H}}} \Psi_{n}^{T}({\bf R}) 
\\   
= & \; e^{-\beta \hat{\mathcal{H}}} \left[  \Phi_{n}^{T}({\bf R})e^{{ \bf i} \phi_{n}({\bf R})} \right] \\ 
= & \; \Phi_{n}^{T}({\bf R},\beta) e^{{ \bf i} \phi_{n}({\bf R})}. \label  {eq:phinau}
\end{align}
Equation (\ref{eq:phinau}) includes by definition all the temperature dependence 
 in $\Phi_n^{T}({\bf R},\beta)~=~\Psi_n^{T}({\bf R},\beta)/e^{{\bf i}\phi_{n} ({\bf R}) } $, denoted as {\it free-amplitude}, since it can be complex,~\cite{stonehenge} with 
 $\Phi_n^{T}({\bf R},0)= \Phi_n^{T}({\bf R})$. The phase 
$\phi_n({\bf  R})$ remains fixed in the interval $[0,\beta)$ as in 
  Ref.~\onlinecite{ortiz93}
 
Following Ref. \onlinecite{stonehenge} we define the quantity 
\begin{align}
\label  {eq:ftau}
f_{n}({\bf R},\beta) =  \Psi_{n}^{T*}({\bf R})\Psi^{T}_{n}({\bf R,\beta})e^{\beta E^X_n},
\end{align}
where $E^X_n$ is a reference energy adjusted numerically to satisfy the
condition  
$\langle \chi^X_n|e^{\hat{J}}|\Psi^{T}_{n}({\beta}) \rangle~=~1$. This reference energy is different from the one commonly used to obtain the ground state. 
In practice, $E^X_n$
depends on the Slater determinant expansion $\chi^{X}_{n}({\bf R})$ used to construct the guiding wave function and
contains the relevant information required to calculate thermodynamical averages.

 Using Eqs~(\ref{eq:H}) and (\ref{eq:U}), one can easily
obtain
\begin{align}
\label  {eq:schrodinger} 
\frac{\partial}{\partial \beta} f_{n}({\bf R},\beta) = 
&   \Psi_{n}^{T*}({\bf R}) \frac{\partial}{\partial \beta} \left[\Psi^{T}_{n}({\bf R,\beta})e^{\beta E^X_n} \right] \nonumber
\\
= &-  \Psi_{n}^{T*}({\bf R}) \left[ (\hat{\mathcal{H}} -E^X_n) \Psi^{T}_{n}({\bf R,\beta})e^{\beta E^X_n} \right] \nonumber \\
= & -\left[E_n^{T}({\bf R },\beta)-E^X_n\right] f_n({\bf R},\beta)
\end{align}
with
\begin{align}
\label  {eq:EL}
E_n^{T}({\bf R},\beta)   & = \frac{\mathcal{\hat {H }} \Psi^T_n({\bf R},\beta)}{\Psi^T_n({\bf R},\beta)}   
\nonumber \\ &    
  = -\frac{1}{2}\!\! 
\sum_j^{N_e}  \frac{\nabla_j^2 \Phi_{n}^{T}({\bf R},\beta)}{\Phi_{n}^{T}({\bf R},\beta)}  \\
 & + 
\frac{1}{2} 
 \sum_j^{N_e}
\left|
{\bf A}_j+{\bf \nabla}_j \phi_{n}({\bf R})
\right|^2
+V({\bf R},\beta) 
\nonumber \\ 
& - {\bf i} \sum_j^{N_e} \left\{
 \frac{{\bf \nabla}_j  \Phi_{n}^{T}({\bf R},\beta)}{ \Phi_{n}^{T}({\bf R},\beta)} \cdot
\left[{\bf A}_j \! + \!{\bf \nabla}_j \phi_{n}({\bf R})\right]
\right .
  \nonumber
\\ 
& \; \; \; \; \; \; \; \; \; \; \; \; \; + 
\left.  
\frac{{\bf \nabla}_j \cdot \left[ {\bf A}_j 
+ {\bf \nabla_j} \phi_{n}({\bf R}) \right]}{2}
\right\}. \nonumber
\end{align}

In order to perform an importance sampling using $\Phi_n^{T}({\bf R})$ as a guiding wave function (as in Refs.~\onlinecite{ceperley80} 
and~\onlinecite{ortiz93}), it is convenient to express 
the kinetic energy in terms of $\Phi_n^{T}({\bf R})$.
The term including  ${\nabla_j^2 \Phi_n^{T}({\bf R},\beta)}/{\Phi_n^{T}({\bf R},\beta)} $ can be 
rewritten using the identity\cite{ceperley80,HLRbook,mfoulkesrmp2001}
\begin{align}
\label  {eq:lapltrn}
 \frac{\nabla^2_j \Phi^{T}_n({\bf R},\beta)}{\Phi_n^{T}({\bf R},\beta)}  = &
\frac{\nabla^2_j f_{n}({\bf R,\beta}) }{f_n({\bf R,\beta})}  
+  \frac{\nabla^2_j \Phi^T_n({\bf R})}{\Phi^T_n({\bf R})}  \nonumber \\
 & - \frac{
 {\bf \nabla}_j \cdot \left[ f_n({\bf R},\beta)
{\bf F}_Q^j 
 \right]
}{f_{n}({\bf R,\beta})} \; ,
\end{align}
where
\begin{align}
\label  {eq:complexgr}
 {\bf F}_Q^j 
  = {\bf \nabla}_j
ln\left|
\Phi_{n}^T({\bf R}) \right|^2 \; .
\end{align}
Note that Eqs. (\ref{eq:lapltrn}) and (\ref{eq:complexgr})  are valid as long as $\Phi^{T}_n({\bf R}) 
\ne 0$ and $f_n({\bf R},\beta) \ne 0$. In practice, any divergence of  $ {\bf F}_Q^j$ at the nodes enforces 
$f_n({{\bf R},\beta})$ to be zero.  Figure \ref{fg:scheme} shows that
complex antisymmetric wave functions can be constructed so that they have nodes only at 
points ${\bf R}$ with ${\bf r}_i = {\bf r}_j$. In this case the nodal error is avoided but errors in the phase introduce
a phase shift\cite{ortiz93,stonehenge} and a complex contribution to $E^L_n({\bf R},\beta)$. However, for complex wave functions without zeros, Eq. (\ref{eq:lapltrn}) is  
always valid, except at the coincidental points (if cusp conditions are not satisfied).  To satisfy Eq. (\ref{eq:lapltrn}) at the coincidental points,  a Jastrow factor is
introduced in Eq. (\ref{eq:trial}). While using complex wave functions involves some complications, the advantage
is that the evolution in imaginary time $\beta$ describes the thermodynamical properties 
with $\beta^{-1} = k_B T$. However, going beyond the fixed-phase approximation~\cite{ortiz93,meyer13} is required to obtain the thermodynamics. In this work the phase is not ``released" in the same sense of Ref. \cite{jones97}, it is only free within the
small subspace. 


Replacing Eq. (\ref{eq:lapltrn})  into Eq (\ref{eq:EL}) and then into  Eq.~(\ref{eq:schrodinger})  one obtains
\begin{align}
\frac{\partial f({\bf R},\beta)}{\partial \beta}
 = &
\frac{1}{2}
 \sum_j^{N_e}
\left\{
\nabla_j^2 f_n({\bf R},\beta) - 
 {\bf \nabla}_j \cdot \left[ f_n({\bf R},\beta)
{\bf F}_Q^j 
 \right]
\right \}
\nonumber \\
& - \left[ E_n^{L}({\bf R},\beta)-E^X_n \right] f_{n}({\bf R},\beta) , 
\label  {eq:far-dmc}
\end{align}
with $E_n^{L}({\bf R},\beta)$  where
\begin{align}
\label  {eq:ELL}
E_n^{L}({\bf R},\beta)  = &   
 -\frac{1}{2}\!\! 
\sum_j^{N_e}  \frac{\nabla_j^2 \Phi_{n}^{T}({\bf R})}{\Phi_{n}^{T}({\bf R})}  \\
 & + 
\frac{1}{2} 
 \sum_j^{N_e}
\left|
{\bf A}_j+{\bf \nabla}_j \phi_{n}({\bf R})
\right|^2
+V({\bf R},\beta) 
\nonumber \\ 
& - {\bf i} \sum_j^{N_e} \left\{
 \frac{{\bf \nabla}_j  \Phi_{n}^{T}({\bf R},\beta)}{ \Phi_{n}^{T}({\bf R},\beta)} \cdot
\left[{\bf A}_j \! + \!{\bf \nabla}_j \phi_{n}({\bf R})\right]
\right .
  \nonumber
\\ 
& \; \; \; \; \; \; \; \; \; \; \; \; \; + 
\left.  
\frac{{\bf \nabla}_j \cdot \left[ {\bf A}_j 
+ {\bf \nabla_j} \phi_{n}({\bf R}) \right]}{2}
\right\}. \nonumber
\end{align}
That is, the local energy  now depends on two guiding  functions  (i) 
$\Phi_{n}^{T}({\bf R})$, and  (ii)   $\Phi_{n}^{T}({\bf R},\beta)$ which is an approximation that
must be obtained and improved for $\beta \ne 0$.

The use of {\it complex valued guiding functions} originates  the gradients  ${\bf \nabla}_j \phi_{n}({\bf R})$ that appear in the local energy  in Eq. (\ref{eq:ELL}). Their
contribution prevents the result from reaching the bosonic solution and enforces an upper bound on
the fermionic ground state.~\cite{ortiz93}  In addition, 
the contribution of  ${\bf \nabla}_j  \Phi_{n}^{T}({\bf R},\beta)$ must
be taken into account in the presence of a magnetic field (see term between the $\{ \}$, when ${\bf A}_j \ne 0$), even when using a real-bosonic guiding wave function with  ${\bf \nabla}_j \phi_{n}({\bf R}) = 0$.

A {\it locality approximation}~\cite{locality}  has been used in the past when a non-local pseudo potential
  is included in $\mathcal{\hat{H}}$ in the potential term $\hat{V}({\bf R})$. It consists in replacing
  $V({\bf R},\beta) = \frac{\hat{V}({\bf R}) \Psi^T_n  ({\bf R},\beta)  }{\Psi^T_n ( {\bf R},\beta)} $ by $V({\bf R},0)$ in Eq.(\ref{eq:EL}). 
Since 
$V({\bf R},\beta) \rightarrow V({\bf R},0) $ 
 for $\beta \rightarrow 0$, we will use the locality approximation
in the first iteration. However, we will improve it in subsequent iterations (see Appendix \ref{ap:beyondloc}). 

A {\it local-time approximation}, analogous to the 
 locality approximation,~\cite{locality}, was introduced in Ref. \onlinecite{stonehenge} to estimate the ratio 
\begin{align}
\label  {eq:gradtrn}
 \frac{{\bf \nabla}_j  \Phi_n^{T}({\bf R},\beta)}{ \Phi_n^{T}({\bf R},\beta)}  = & 
 \frac{{\bf \nabla}_j  \Phi_n^{T}({\bf R})}{ \Phi_n^{T}({\bf R})} +
 {\bf \nabla}_j \ln\left[    \frac{ \Phi_n^{T}({\bf R},\beta)}{ \Phi_n^{T}({\bf R})}   \right].
\end{align}
One can neglect the last term in Eq. (\ref{eq:gradtrn})  for $\beta \rightarrow 0 $ where $ { \Phi_n^{T}({\bf R},\beta)}/{ \Phi_n^{T}({\bf R})} \rightarrow 1 $.  In the present work we will use the local-time approximation only in the first iteration. For subsequent iterations we improve the evaluation of Eq. (\ref{eq:gradtrn})  by using the sampling  of the dependence in 
$\beta$ of $\Phi_n^{T}({\bf R},\beta)$ obtained in the previous iteration.

The locality and local-time approximations have little impact in optimization methods 
that focus on eigenstates because the dependence on $\beta$ of $\Psi^{T}_n({\bf R,\beta})$ is minimized when the optimization
progresses as 
$\Psi^{T}_n({\bf R}) \rightarrow \Psi_n({\bf R})$. Going beyond these approximations is required, however,  to
circumvent the nodes of the eigenstates with  complex wave functions.  Fortunately, optimization in the small subspace allows an easy sampling of the $\beta$ dependence. The approach is exact if the big basis is large enough and if enough statistical data is 
collected~\cite{fn:fixedphase}  as $\ell \rightarrow \infty$.

{\bf Circumventing the nodes} with complex wave functions is necessary in this case because, in standard DMC calculations using real-valued wave functions 
with nodes, if any walker crosses a 
node, it is either 
killed\cite{ceperley80} or the move is rejected.~\cite{umrigar93} This introduces an 
artificial divergent 
potential at the nodal surface, which adds a kink at the node (a step for the rejection case). Since there is 
a one-to-one correspondence between energy of one eigenstate and its nodes,~\cite{rosetta} eigenstates with 
different energies must have different nodes. As a consequence, two real wave wave functions that approach 
different eigenstates introduce different nodal potentials.  Since the fixed-node Hamiltonian is 
different for different eigenstates, and affect the dynamics at the node in the evolution in imaginary time, the $\beta$ dependence obtained using the fixed-node approximation will not describe
 the thermodynamics even if the exact nodes of the ground state are provided. 

\subsection{Differences with other DMC-like projection methods}
The  implementation of this method follows essentially the same approach developed for DMC or SHDMC, 
with some key numerical changes. 

Equation (\ref{eq:far-dmc}) is identical to Eq. (1) in Ref. \onlinecite{ceperley80} except for the local energy, which now has an explicit dependence in $\beta$. Unlike Eq. (13) in Ref. \onlinecite{stonehenge},  Eq. (\ref{eq:far-dmc})  is now valid for $\beta \ge 0$.
As in Ref  \onlinecite{ceperley80}, Eq (\ref{eq:far-dmc}) describes the evolution of an ensemble.  
Each member of the ensemble of configurations ${\bf R}_i$ (walker) undergoes (i)
a random diffusion and (ii) drifting
by the quantum force $ \sum_j {\bf F}_Q^j $
(which depends only on $\Phi_T({\bf R})$ and not on the
phase). Following Ref. \onlinecite{stonehenge}, (iii)
each walker carries a complex phase factor. In a nonbranching algorithm, the
complex weight of the walkers is multiplied by $\exp\{-\left[ E^L_n({\bf  R},\beta)-E^{X}_n \right]\delta\beta\}$ at every 
diffusion step. In contrast with Ref. \onlinecite{jones97}, the phase factor of the walkers starts in 1 and evolves towards 
{\it the difference} between the guiding phase and the phase of $\Psi({\bf R},\beta)$, while in the release phase approach the initial 
phase of each walker depends on the initial positions of the walkers ${\bf R}_i$ but  remains constant in $\beta$.  

 If the $\chi_n^X ({\bf R})$ are  linear combinations of 
antisymmetric functions, with arbitrary complex coefficients, the $\Phi_n^T ({\bf R})$ in Eq. (\ref{eq:ELL}) do not have nodal surfaces (see Fig. \ref{fg:scheme}). Therefore,
 ${\bf F}_q$ is not divergent, but at the coincidental points.  
 
{\it All walkers must add the same  inverse temperature $\beta$ after $k$ steps.}
 The  standard time-step correction to minimize time step errors at the nodes [Eq.~(33) in
Ref.~\onlinecite{umrigar93}] is not strictly necessary since the divergences in ${\bf F}_q$ are removed.  If one introduces it, one must readjust $\delta \beta$ during the time evolution 
so that all the walkers add up to the same  $\beta$. For the same reason, the standard accept/reject scheme that 
enforces detailed balance\cite{HLRbook} is modified: other moves are retried after rejection until a move 
is accepted. 
\subsection{Calculation of the partition function}

This section shows that the partition function  can be obtained as  $Z(\beta)\simeq \sum_{n=0}^{M_S-1}  e^{-\beta E^{S}_{n}}  $ where 
the $E^{S}_n$ are reference energies instead of eigen energies.  

Numerically, it is convenient to start the calculation with a distribution of walkers proportional to 
$f_n({\bf R},0)=|\Phi^T_{n}({\bf R})|^2$. 
In the importance sampling approach~\cite{ceperley80} setting the second line of Eq. (\ref{eq:far-dmc}) equal to zero, provides an equilibrium distribution proportional to $ |\Phi^T_n({\bf R})|^2 $. 

As in the DMC and SHDMC methods, the evolution in inverse temperature $\beta$ is discretized 
into $k$ finite steps $\delta\beta=\beta/k$. Following the SHDMC approach\cite{keystone,rockandroll,rollingstones,stonehenge}
the weighted distribution of the walkers can be written as
\begin{align}\label  {eq:evoltau}
  f_{n}({\bf R},k \delta\beta ) 
                & =  \lim_{N_c \rightarrow \infty} \frac{1}{N_c} \sum_{i=1}^{N_c}
   W_i \delta \left({\bf R-R}_i \right). 
\end{align}
In Eq.~(\ref{eq:evoltau}), ${\bf R}_i$ corresponds to the position
of the walker $i$ , and  $N_c$ is the number of equilibrated
configurations. The {\it complex} weights $W_i$ are given by
\begin{align}
\label  {eq:weights}
W_i = e^{-\left[
E_i^k-E^{X}_{n}
\right]  k \delta\beta}
\end{align}
with 
\begin{align}
  E_i^k = \frac{1}{k}\!\sum_{j=0}^{k-1} E^L_n({\bf R}_i^{-j}),
\end{align} 
Where $k$ is a number of steps and $E_L({\bf R}_i^{-j})$ is the previous value of the local energy obtained $j$ 
 steps $\delta\beta$ earlier for the walker $i$.
 
 The  evolution in inverse temperature $\beta$ of the guiding wave function $\Psi_n^{T}({\bf R},\beta)$ 
 can be written, without loss of generality, as  
 \begin{align}
 \label  {eq:beyondloc}
 \Psi_n^{T}({\bf R},\beta) =
 e^{-\beta E^{X}_{n}} e^{-J({\bf R})} \left[\chi^{X}_n({\bf R})+\delta\chi^{X}_n({\bf R},\beta) \right]. 
 \end{align}
 That is, the product of an average decay factor  $e^{-\beta E^{X}_{n}} $ times the Slater determinant part. The Slater part is given by the
 the one at $\beta=0$ plus an orthogonal displacement $\delta\chi^{X}_n({\bf R},\beta)$. 
 The $X$ in term $\delta\chi^{X}_n({\bf R},\beta)$  denotes the explicit dependence on $\chi^{X}_n({\bf R})$. At least one overlap $\langle \chi^{X}_m| \delta\chi^{X}_n,\beta \rangle $ 
 must be different from zero for $n \ne m$ , if  $\Psi^T_n({\bf R})  $ is not an eigenstate. 
 
 Using Eq. (\ref{eq:beyondloc}) we can correct equation (\ref{eq:ELL})  beyond the locality and local-time approximations. 
 The displacement $\delta\chi^{X}_n({\bf R},\beta)$    can be
 sampled from the DMC run as follows:  
From Eqs.~(\ref{eq:U}),  (\ref{eq:ftau}) and  (\ref{eq:evoltau}), one can
 obtain
\begin{align}
\label  {eq:nextevol}
e^{-J({\bf R})} \delta \chi^{X}_n({\bf R},\beta)   = & \langle {\bf R} | \left [
 e^{\beta E^X_n}|\Psi^{T}_n (\beta) \rangle  - |\Psi^{T}_n \rangle \right ] \nonumber \\
   =& \; \langle {\bf R} | \left[e^{-\beta (\hat{\mathcal{H}}-E^{X}_{n})} -1 \right ]e^{-\hat{J}}|\chi^{X}_n \rangle \; \\
  = & \;  \frac{ \left[ f_n({\bf R}, \beta) -f_n({\bf R},0) \right]} 
{ \Psi^{T *}_n({\bf R}) } \\
   \label  {eq:deltaevol}
\simeq & \frac{1}{N_c} \sum_{i=1}^{N_c} e^{ J({\bf
    R})} \frac {[W_i-1] } 
{ \chi^{X*}_n ({\bf    R})} \delta({\bf R}-{\bf R}_i)
. 
\end{align}
Within the subspace spanned by the basis $\{\chi^{S}_n({\bf R}) \}$, the identity operator $\hat{E}$ is given by
\begin{align} 
\label  {eq:deltaexp}
\langle {\bf R}^{\prime}| \hat{E} |  {\bf R} \rangle & = 
\sum_{m=0}^{M_S-1}
 e^{-J({\bf R^\prime})}\chi^{S}_m({\bf R^{\prime}}) \chi^{S*}_m({\bf R}) e^{J({\bf R})}
.
\end{align}

Applying Eq.~(\ref{eq:deltaexp}) to both sides of Eq.~(\ref{eq:deltaevol}), 
and integrating over ${\bf R}$, 
one can easily obtain an expression of the diffusion displacement   within the basis $ \{ | \chi^{S}_n \rangle \} $ as
\begin{align}
\label  {eq:trialwf}
\delta\chi^{X}_n({\bf R},\beta ) & = 
\sum_{m=0}^{M_S-1} \lambda_n^m(\beta)  \chi^{S}_m({\bf R}) 
\end{align}
with 
\begin{align} 
\label  {eq:deltalambda}
\\ \nonumber &
\lambda_n^m(\beta)  = \frac{1}{N_c} \sum_{i=1}^{N_c} e^{ 2J({\bf
    R}_i)} \frac {\chi^{S *}_m ({\bf R}_i)} 
{ \chi^{X*}_n ({\bf    R}_i)} [W_i-1] 
\end{align}
 where $ N_c= \sum_{i=1}^{N_c} W_i $.

Using $X=S$ in Eqs. (\ref{eq:evoltau}) -(\ref{eq:beyondloc}) one can prove that 
\begin{align}
\label  {eq:JUJ}
\langle \chi^{S}_n | \mathcal{\hat{U}}| \chi^{S}_m \rangle = & \int {\bf dR} e^{J({\bf R})}\chi^{S*}_n ({\bf R})\Psi_m^T ({\bf R},\beta) \nonumber \\
= & e^{-\beta E^{S}_{n}} \left[ \delta_{n,m} +                   
 \int {\bf dR} \chi^{S*}_n ({\bf R})\delta\chi^{S}_m ({\bf R},\beta) 
\right]  \nonumber \\
= & e^{-\beta E^{S}_{n}} ( \delta_{n,m}+ \lambda_n^m ) 
\end{align}
with $\mathcal{\hat{U}}$ having the structure of the transcorrelated method~\cite{ten-no02}
\begin{align}
\mathcal{\hat{U}}= & e^{\hat J} e^{-\beta \mathcal{\hat{H}}} e^{- \hat J}.
\end{align}
We use condition $ \lambda_n^n = 0 $ [See Eq (\ref{eq:deltalambda})] to determine the value of  $E^{S}_{n}$. 
In practice, we adjust the reference energy of the guiding functions every iteration as  $e^{-\beta E^{X,\ell+1}_{n}}= e^{-\beta E^{X,\ell}_{n}} ( \delta_{n,m}+ \lambda_n^m )$.

Since $tr( e^{-\beta \mathcal{\hat{H}}}) = tr(\mathcal{\hat{U}} )  $, the contribution to the
Helmholtz free energy of the small subspace is given by
\begin{align}
\label  {eq:F}
F(\beta)=- \frac{1}{\beta} \ln{\left[\sum_{n=0}^{M_S-1}  e^{-\beta E^{S}_{n}}  \right]},
\end{align}
where the expression inside the brackets is the partition function $Z(\beta)$. 

In general for an arbitrary  guiding function $\Psi_m^T ({\bf R}) $, the variance will grow with $M_S$. An energy span 
larger than $\beta^{-1}$ must be retained in the basis to calculate
thermodynamical properties. Arbitrary trial wave functions spanned by this space might have significant variance 
in the walkers weights. To reduce the variance we
use guiding functions that are approximately linear combinations of a pair of neighboring eigenstates.   

When using guiding functions that are different from the small basis functions, the  contributions to the trace of the density 
matrix in the small basis can be obtained with 
\begin{align}
 e^{-\beta E^{S}_n} = \sum_i 
 \left[
 |\langle \chi^{X}_i  |\chi^{S}_n \rangle|^2+\langle \chi^{X}_i  |\chi^{S}_n \rangle \langle \chi^{S}_n  |\delta\chi^{X}_i \rangle
 \right] e^{-\beta E^X_i} \; .
 \end{align}
The details of the derivation are in Appendix~\ref{ap:eigenstates}.  

In a recent paper, Mazzola, Zen and Sorella\cite{mazzola12} proved that   
\begin{align}
\label{eq:mazzola}
\langle 
\Psi^T_n | e^{-\beta \hat{\mathcal{H}}} |
\Psi^T_n 
 \rangle  \ge 
 e^{- \beta
 \langle
 \Psi^T_n | \hat{\mathcal{H}} |
\Psi^T_n ,
\rangle 
}.
\end{align}
 Ref. \onlinecite{mazzola12} used the righthand side of  
 Eq. (\ref{eq:mazzola}) to approximate the free energy obtaining a lower bound for $F(\beta)$.  Reference \onlinecite{mazzola12} can be considered a VMC approach to the evaluation of the free energy. That approximation becomes exact if all the $|\Psi^T_n \rangle$ are 
 eigenstates of $\hat{\mathcal{H}}$. However, that method is very poor for an arbitrary random guiding function. In the present approach, we 
 go beyond Ref. \onlinecite{mazzola12} by evaluating the lefthand side of Eq. (\ref{eq:mazzola}) directly using DMC.

In many situations, the excitations of a mean field method based on approximations DFT might be good enough to obtain the low energy thermodynamical properties using Eq. (\ref{eq:F}). If that were the case, at least two DMC runs for each function of the basis are required. One to obtain the $\beta$ dependence and a second to evaluate the reference energies beyond the local-time approximation. However, in the so-called
highly correlated materials, usual approximations of DFT fail to describe the low energy physics. In those cases a method that could optimize the basis
is more important. That method is described in the following subsections.

\subsection{The first iteration: Construction of the first small basis  $\{ |\chi^{S,1}_n \rangle \}$}
\label{ssc:initialization}

While the present approach will optimize the basis from any starting basis set, the calculation will be more efficient starting from a good basis.  
A procedure to generate a good starting set is described here. 

The only restriction for the small basis $\{ |\chi^{S}_n \rangle \}$ is to avoid the nodes associated with real wave functions.  In this work we choose
the initial basis  with a Lanczos-like procedure combined with the SHDMC approach. 

The big subspace basis set ${| m \rangle}$ is constructed by   symmetry constrained functions (linear combinations of Slater determinants with the
same symmetry of the ground state) ordered with increasing mean field energy.

We choose the first basis function of the small subspace to be
\begin{align}
\label  {eq:ini0}
| \Psi_0^{S} \rangle  = \frac{1}{\sqrt{2}}(|0 \rangle + {\bf i} | 1 \rangle )
\end{align}
being $|0 \rangle$ and $|1 \rangle$the ground and first excited states of a non-interacting solution 
of the system. 

Using Eq. (\ref{eq:ini0}) as guiding function in Eqs. (\ref{eq:nextevol})-(\ref{eq:deltaevol}) and replacing $\chi^{S}_m({\bf R})$ by $\langle {\bf R} | m \rangle$ in 
Eq. (\ref{eq:deltaexp}), but not on $\Psi^T_n({\bf R})$,  we obtain an expression 
similar to Eq. (\ref{eq:trialwf}) 
\begin{align}
\label  {eq:projection}
\delta \tilde{\chi}_n^{S} ({\bf R}) =\langle {\bf R} | \delta \tilde{\chi}^{S}_n \rangle = \tilde{\sum}_{m=0} c_n^m \langle {\bf R} | m \rangle.
\end{align}
The tilde in  $\delta \tilde{\chi}_n^{S} ({\bf R}) $ means that the expansion 
is in the big basis $\{ |m \rangle \}$  with 
\begin{align} 
\label  {eq:cnm}
\\ \nonumber &
c_n^m(\beta)  = \frac{1}{N_c} \sum_{i=1}^{N_c} 
e^{2 J({\bf    R}_i)} \frac {\langle m |{\bf R}_i \rangle} 
{ \chi^{S*}_n ({\bf    R}_i)} [W_i-1] .
\end{align}

The symbol $\tilde{\sum}$  in Eq. (\ref{eq:projection}) 
means that the sum is restricted to the coefficients $c_n^m$ with an error bar smaller than $25\%$ of the absolute value 
(this is the standard recipe of the SHDMC algorithm~\cite{keystone}). 

We define the next basis function $|\chi^{S}_{n+1} \rangle $ recursively as
\begin{align}
\label  {eq:inin}
|\chi^{S}_{n+1} \rangle =& \frac{1}{\mathcal{N}_{n+1}} \hat{P}_{n+1}  |\delta\tilde{\chi}^{S}_{n} \rangle  \text{ with } \\
\label  {eq:Pn}
\hat{P}_n=& 1-\sum_{m=0}^{n-1} | \chi^{S}_m \rangle  \langle \chi^{S}_m |
\end{align}
 where $\mathcal{N}_{n+1}$ is a normalization constant. Equations (\ref{eq:Pn}) and (\ref{eq:inin})  mean that $|\chi^{S}_{n} \rangle $ is
the projection of the displacement $|\delta\tilde{\chi}^{S}_{n+1} \rangle$ orthogonal to the subspace spanned by the $n$ basis
functions found previously. 

One repeats this procedure until a basis of $M_S$ functions $\{ |\chi^{S}_n \rangle \}$ is constructed. This Lanczos-like procedure grants that the initial small subspace basis times the 
Jastrow factor has a large projection
onto the lowest-energy eigenstates of $\mathcal{\hat{H}}$ or the largest eigenstates of $e^{- \beta \mathcal{\hat{H}}}$.

Since the evolution in inverse temperature $\beta$ is not known during the initialization step, we use
the local-time approximation discussed in the previous section. However, once a basis is generated, 
 we can go beyond the local-time approximation in successive iterations. Note that by construction any 
$\chi^{S}_n({\bf R},\beta)$ can be approximated as a linear combination of the basis function $\chi^{S}_m({\bf R})$ 
with $m<n+1$, since the finite temperature projection of one wave function of the basis into the other was used to construct the small basis.
The details on how to approximate the evolution in $\beta$ are in Appendix \ref{ap:beyondloc}. 

\subsection{Systematic improvement of the small basis  $\{ |\chi^{S}_n \rangle \}$} 

One of the goals of this work is to obtain a much smaller basis of  $M_S$  functions $\{|\chi^{S}_n \rangle \}$ than $\{| m \rangle\}$,
the big set of $N_B$ basis functions. 
The small basis $\{  |\chi^{S}_n \rangle\}$ should retain the
lowest energy physics of $\mathcal{\hat{H}}$ and  $e^{-\beta \mathcal{\hat{H}}}$. While for some purposes (e.g. the 
ground state calculation),
the initial basis set described in the previous section might be enough, in this subsection 
we describe how to further optimize  the small
basis  so it better describes the low-energy excitations of  $\mathcal{\hat{H}}$ and thermodynamical properties. 

Note that 
$\chi^{S}_n({\bf R},\beta)=e^{\hat J} e^{-\beta \mathcal{\hat{H}}} e^{- \hat J} \chi^{S}_n({\bf R})$ will converge 
to the antisymmetric part of the ground state wave-function as $\beta \rightarrow \infty$. In order to avoid every state in the basis 
 collapsing to the same function we (i) remove the projection into the other states of the basis, (ii)  add the diffusion
displacement   orthogonal to the small subspace, and (iii) perform a  GramÐSchmidt  orthogonalization as follows:
\begin{align}
\label  {eq:improven}
|\chi^{S,\ell+1}_0 \rangle = & |\chi^{S,\ell}_0 \rangle + |\delta\tilde{\chi}_0^{S} \rangle -|\delta\chi^{S}_0(\beta) \rangle \\
|\chi^{S,\ell+1}_n \rangle = & \hat{P}^{\ell+1}_n \left[ 
|\chi^{S,\ell}_n \rangle + |\delta\tilde{\chi}_n^{S} \rangle -|\delta\chi^{S}_1(\beta) \rangle \right] \;\; n>0, \nonumber
\end{align}
with $\hat{P}^{\ell+1}_n $ given by Eq. (\ref{eq:Pn}) replacing $\chi^{S}_m $ by $\chi^{S,\ell+1}_m $.
Note in Eq. (\ref{eq:improven}) that $ |\delta\tilde{\chi}_n^{S} \rangle $ given by Eq. (\ref{eq:projection}) is a direct projection of the diffusion displacement  
into the big basis of SCFs $\{ |m \rangle \}$, whereas $|\delta\chi^{S}_n (\beta)\rangle $ given by Eq. (\ref{eq:trialwf}) is an indirect projection (since the  $|\delta\chi^{S}_n (\beta)\rangle $ are projected into the small basis $\{ |\chi^{S}_n \rangle \}$ which in turn are
linear expansions of functions that belong to $\{ |m\rangle \}$). The difference $ |\Delta\tilde{\chi}^{S}_n \rangle= |\delta\tilde{\chi}^{S}_n \rangle -|\delta\chi^{S}_n(\beta) \rangle$ is by construction orthogonal to the small subspace. Accordingly, it describes the decay of the small subspace basis into the eigenstates with the
lowests energies.

\section{Algorithm}
\label{sc:algorithm}

The goal of this algorithm is to optimize a minimal basis $\{ |\chi^{S}_n \rangle \}$ to span  the lowest energy excitations of $\mathcal{\hat H}$ 
(equivalently the  eigenstates of $e^{-\beta \mathcal{\hat H}}$ with the largest eigenvalues). That basis can be used
to calculate finite temperature expectation values of thermodynamical properties and accelerate the calculation of the ground and lower excited states. In this section we summarize how the theory described in detail earlier can be 
implemented. 

{\bf Initialization:} the small basis $\{ |\chi^{S}_n \rangle \}$, a set of orthogonal linear combinations of many-body  functions $|m \rangle$
 is constructed using the procedure described in  \ref{ssc:initialization}. Once this procedure is concluded, 
a set of linearly 
independent guiding  functions  $e^{-\hat{J}} | \chi^{V}_n \rangle$ is constructed as linear combinations of
 pairs of approximated eigenstates of 
$ \mathcal{\hat{U}}$.

 Subsequently, we can use for the evaluation of the 
local energy $E^L_n({\bf R},\beta)$ an approximate dependence of the guiding functions in $\beta$ given by Eq. (\ref{eq:pairsbeta}).

{\bf Basis update iteration:} each iteration $\ell$ is composed by (1) a parallel diffusion of intermediate 
functions  $| \chi^{V}_n \rangle$. (2)  A linear transformation to obtain  
 $|\delta \tilde{\chi}^{S}_n\rangle$ and   $|\delta \chi^{S}_n \rangle$.  (3)
Recalculation of $\mathcal{\hat{U}}$ in the small basis. (4) Update of the small basis for the
next iteration $| \tilde{\chi}^{S,\ell+1}_n\rangle$. (5)
Update of the intermediate functions   $| \chi^{V,\ell+1}_n \rangle$. (6) Finally, the algorithm decides 
to increase the number of samples in the next iteration or not.  These six steps are repeated recursively. 

The individual steps  of the iteration are described below in more detail:  

(1) Parallel diffusion: Each displacement  
 \begin{align}
 \label  {eq:dsv}
 |\delta\chi^{V}_n \rangle=e^{\hat{J}}\left[e^{-\beta \mathcal{\hat{H}}}  -1\right]e^{-\hat{J}}| \chi^{V}_n \rangle
 \end{align} 
 is projected 
into the small basis $\{ |\chi^{S}_n \rangle \}$ and the big basis $\{| m \rangle \}$ using Eqs. (\ref{eq:deltalambda}) and Eq. (\ref{eq:cnm}) respectively
replacing $\chi^{X}_n ( {\bf R}) $ by $\chi^{V}_n ( {\bf R}) $. 

Each diffusion contains $M_b$ sampling subblocks. 
For each sampling subblock, uncorrelated walker positions are generated from 
the previous one with a VMC algorithm. 
Next, DMC is run 
for $k$ steps with a shorter time step $\delta \beta$. 
The coefficients of  $|\delta \tilde{\chi}^{V}_n\rangle$ and   $|\delta \chi^{V}_n \rangle$
are sampled at the end of 
each subblock using Eqs. (\ref{eq:deltalambda}) and (\ref{eq:cnm}).  Statistical data is collected for
$M_b$ subblocks for each parallel diffusion before an update of the small basis. 

(2)  Linear transformation: 
The $|\delta \tilde{\chi}^{S}_n\rangle$ and   $|\delta \chi^{S}_n \rangle$ can be constructed in terms of 
$|\delta \tilde{\chi}^{V}_n\rangle$ and   $|\delta \chi^{V}_n \rangle$ using 
Eq. (\ref{eq:conversion}) replacing the super index $X$ by $V$.

(3) Calculation of $\mathcal{\hat{U}}$,     $\{|\Psi^{U,\ell+1}_n \rangle \}$, and  $\{|\Psi^{V,\ell+1}_n \rangle \}$  :
A matrix representation of $\mathcal{\hat{U}}$ is obtained in the small subspace $\{|\Psi^{U,\ell}_n \rangle \}$,
using Eqs. (\ref{eq:deltalambda}), (\ref{eq:JUJ}) and (\ref{eq:conversion}). The left and right eigenvectors are obtained by diagonalizing $\mathcal{\hat{U}}$.  
 
(4) Update of the small Basis:  Equation (\ref{eq:improven}) is used to improve the small basis.

(5) We perform the correspondence  
$\chi^{S,\ell}_n \rightarrow \chi^{S,\ell+1}_n$ (see Appendix \ref{ap:beyondloc} ) and construct the basis 
$\{|\chi^{U,\ell+1}_n \rangle \}$, and  $\{|\chi^{V,\ell+1}_n \rangle \}$   with the coefficients of the eigenvectors of
 $\mathcal{\hat{U}}$ in the iteration $\ell$.

(6) Updating $M_b$: At first, the number of sampling subblocks $M_b$
is set to a small number (e.g. $M_b = 3$).  
When the noise is dominant    
$\sum_n \langle \Delta\tilde{\chi}^{S,\ell+1}_n | \Delta\tilde{\chi}^{S,\ell}_n \rangle \le 0$, we increase 
 $M_b$ by a factor larger than 1.  As a result, the
total number of configurations $N_c$ sampled increases as the
iteration $\ell$ increases and the statistical error is reduced. Hence, as the statistics are improved 
the number of basis functions retained in the expansion Eq.~(\ref{eq:deltaexp}) increases over
time.  

(7) We use Eq. (\ref{eq:F}) to calculate thermodynamical properties.  

\section{Results in model calculations}
\label{sc:results}
This section describes the results obtained  for a model system with an applied magnetic field.
The results are compared with configuration interaction (CI)
calculations in the same model used in Refs. \onlinecite{rosetta,keystone,rockandroll} and \onlinecite{stonehenge}.
The lowest-energy eigenstates were found for two polarized electrons ($J=1$)
moving in a two-dimensional square with a side length $1$ and a
repulsive interaction potential of the form $\hat{V}({\bf r},{\bf
  r^{\prime}}) = 8 \pi^2 \gamma \cos{[\alpha
  \pi(x-x^{\prime})]}\cos{[\alpha \pi(y-y^{\prime})]}$ with $\alpha=
1/ \pi$ and $\gamma = 4$.  

The main advantage of the model is that fully converged CI calculations can be performed which are nearly analytical. 
In order to perform the CI calculations
the many-body wave function of the small basis $\{\chi^S_n \rangle\}$ are spanned in a basis of 
functions $\{ | m \rangle \}$ that are eigenstates of the noninteracting
system. They are linear
combinations of functions of the form $\prod_{\nu} \sin(m_{\nu} \pi
x_{\nu})$ with $m_{\nu} \le 7$.  Converged CI calculations were
performed to obtain a nearly exact expression of the lowest energy
states of the system $| \Psi_n \rangle = \sum_m a_m^n  | m \rangle$.
The matrix elements involving the magnetic vector potential ${\bf A}$
(in the symmetric gauge) were calculated analytically.  The result of the
CI calculations were used to evaluate the partition functions and to quantify the 
convergence of the basis.

The same basis used to construct the CI Hamiltonian is used as the {\it big basis} to test our finite temperature version of SHDMC. 
All the calculations reported are with $J({\bf R}) = 0$, which increases the statistically noise, makes the test more difficult and facilitates the comparison with the CI results. The results presented here are a proof of principle 
on the validity of the algorithm, which is necessary before requesting and using the massive amount
of computing time required for realistic finite temperature calculations in solids. While clearly a demonstration in a realistic system is required in the future,  a comparison with an exact model is the first essential step  to validate   
 the scheme. This includes not only the value obtained for
 the partition function but also a detailed analysis of the convergence of the basis.

\begin{figure}
\includegraphics[width=1.00\linewidth,clip=true]{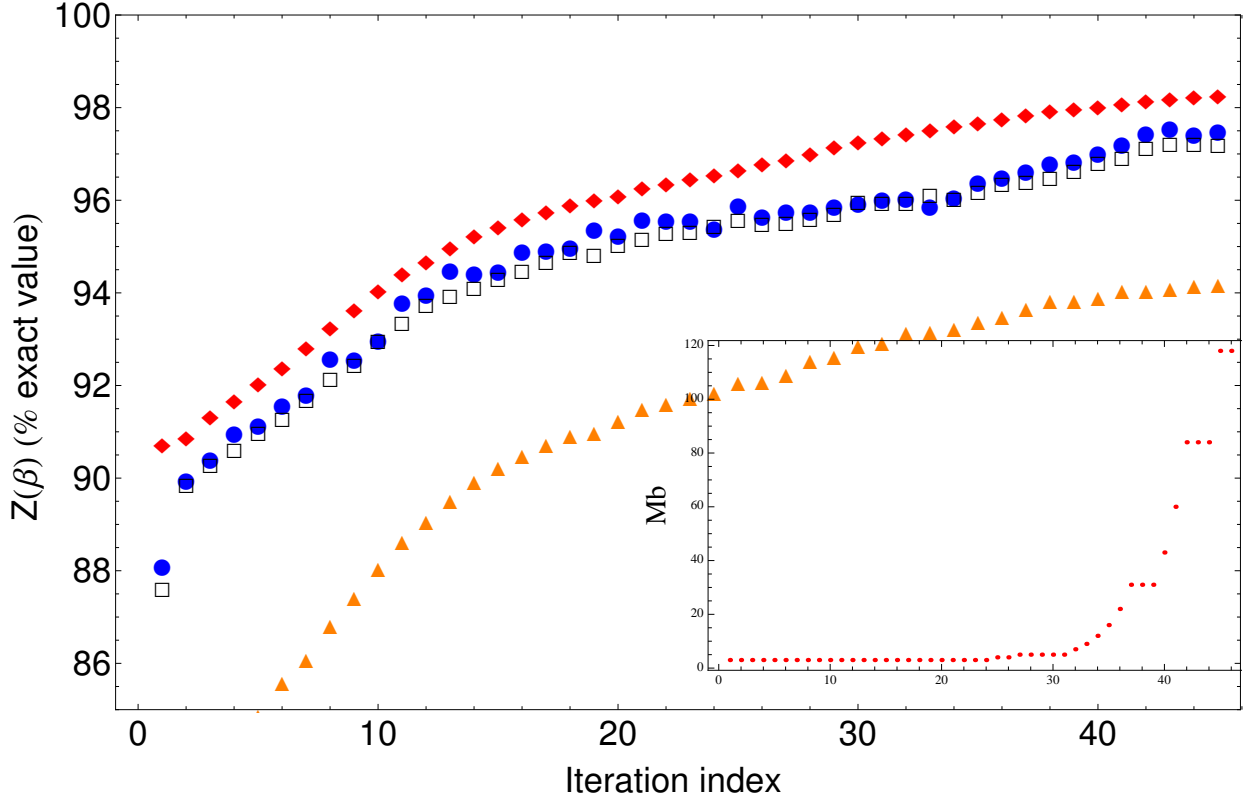}
\caption{
\label{fg:trace} (Color online) Evolution of the trace of $\mathcal{\hat{U}}= e^{\hat{J}} e^{-\beta \mathcal{\hat H}}e^{-\hat{J}}$ in the
subspace defined by the small basis $\{ | \chi^{S,\ell}_n \rangle \}$ as a function of the iteration index $\ell$ for $M_S=20$ relative
to the truncated trace obtained with full CI eigenvalues. Note
that the scale starts from 86\% of the full CI value. The blue circles are results obtained with SHDMC using Eq. (\ref{eq:F}).  
The red rhombi denote the exact evaluation of $\mathcal{\hat{U}}= e^{\hat{J}} e^{-\beta \mathcal{\hat H}}e^{-\hat{J}}$ in the small basis, using
full CI data. The
empty squares (up triangles) were obtained by evaluating $\langle
 \Psi^T_n | \hat{\mathcal{H}} |
\Psi^T_n ,
\rangle $ for linear combination of eigenstate pairs 
(random rotation within the small subspace) in the full CI basis while $Z(\beta)$ was obtained using an lower bound approximation.~\cite{mazzola12}  The inset shows the computational cost, the number of sampling subblocks $M_b$, as a function of the iteration. }
\end{figure}

In the absence of magnetic fields there are two degenerate solutions: one that transforms line $x$, and
the other that transform like $y$.  This degeneracy is broken with a magnetic field.
The eigenstates transform like $x+{\bf i} y$ and $x-{\bf i} y$. 
Figure \ref{fg:trace} shows the evolution of 
$Z_+(\beta)=tr_+(\mathcal{\hat{U}})$ of the  model system with $M_S=20$. 
The subindex ``$+$" in $Z_+$ and $tr_+$ means 
that the results of 
Fig. (\ref{fg:trace}) were obtained considering only the  subspace of the Hamiltonian that transforms like $x+{\bf i} y$. The calculation of 
thermodynamical properties requires, however, the inclusion of all possible symmetries 
of the wave function, which implies that $Z_-$, the trace in a small basis that transforms as $x-{\bf i} y$, should
also be added. In order to calculate $Z_+(\beta)$ we have defined the zero of energy to be the ground state
of the CI.   The calculations were run using $\delta \beta=0.00002$ and
$\beta=0.004$. We have used a magnetic field of $B=0.6283$. 

Figure \ref{fg:trace} shows the value of $Z_+(\beta )$, calculated with different methods, relative to the
full CI value obtained with the $M_S=20$ lowest 
eigenvalues. The blue cycles were obtained with SHDMC using Eq. (\ref{eq:F}). The red rhombi were obtained 
by evaluating  $Z_+(\beta)=\sum_{n,m} |\langle \Psi^T_m | \Psi_n \rangle |^2 e^{-\beta E_n } $, being the $| \Psi_n \rangle$ and $E_n$ the 
eigenvectors and eigenvalues of the full CI. 
The empty squares were obtained as 
$Z_+(\beta) \simeq \sum_n  e^{- \beta \langle | \Psi^T_n | \hat{\mathcal{H}} | \Psi^T_n \rangle }$ being the $| \Psi^T_n \rangle$ linear 
combinations of pairs of approximated eigenstates obtained with SHDMC.  Therefore, the squares correspond to the result that one
would had obtained using the approximation of Ref. \onlinecite{mazzola12} in Eq. (\ref{eq:mazzola}) 
for a very good set of functions. Finally,
the up triangles mark the result obtained with 
$Z_+(\beta) \simeq \sum_n  e^{- \beta \langle | \Psi^T_n | \hat{R^{-1}}  \hat{\mathcal{H}}  \hat{R} | \Psi^T_n \rangle }$, being $\hat{R}$
a random rotation defined in the small subspace.  

Figure \ref{fg:trace} shows that as the iteration $\ell$ increases, the $Z_+(\beta )$ obtained with all methods 
increases. Note, that the scale of the $y$ axis does not
start from zero. The initialization scheme already produces a
basis that retains 90 \% of the exact value of the truncated partition function.
Similar to previous SHDMC methods, the present generalization optimizes the small basis overlap, not their average energy. The trace 
 of $\mathcal{\hat{U}}$ increases indirectly as the small basis approaches to the subspace of the eigenvectors with lowest energy.
 Note that the SHDMC results are within $\sim 1.5$\%  below the values obtained analytically by projection into the CI data (red rhombi).
  This difference is due to the remaining
 errors in the evolution phase which neglects the projection orthogonal to the small subspace. The method used 
 in Ref. \onlinecite{mazzola12} applied to approximated pairs of eigenstates (empty squares) gives results only slightly below the SHDMC values, because the energy difference between eigenstates is much smaller that  $\beta^{-1}$. However, a random rotation of the basis that spans the 
 same subspace (up triangles) would had produced a significantly worse result. This shows that the approach described
 in Ref. 12 is significantly worse if each element of the basis does not have a large projection on each eigenstate. 
 
  Note
 in the inset of Fig. \ref{fg:trace} that the number of sampling blocks remains very low ($M_b=3$) for the first 25 iterations and starts increasing 
 when the noise becomes dominant around $\ell = 32$. If one considers only the first 32 iterations a significant improvement of the partition function is achieved with little computational cost as
 compared with the calculation of individual eigenstates. 

\begin{figure}
 \includegraphics[width=1.00\linewidth,clip=true]{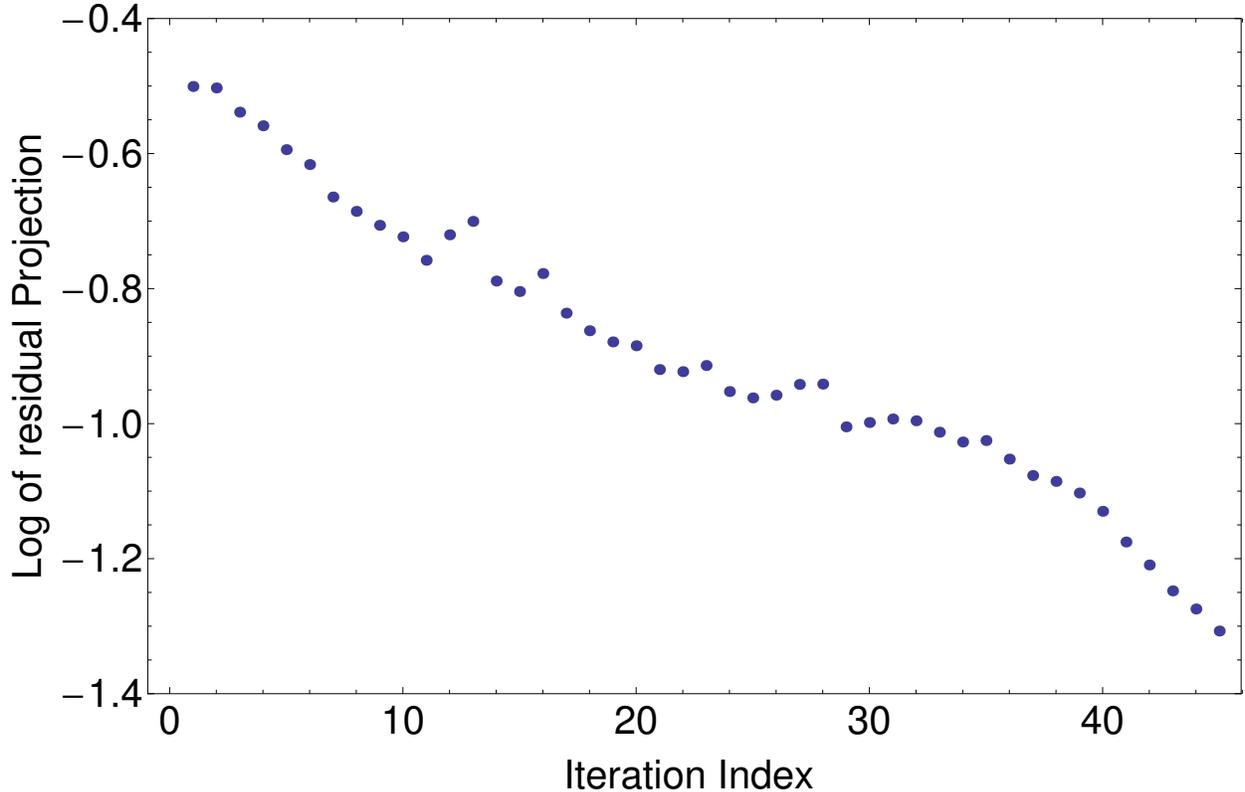}
 \caption{
 \label{fg:determinant} Logarithm of the residual subspace projection $LR_{sp}$ 
 [see Eq. \ref{eq:determinant}] of the small basis $\{ | \chi^{S,\ell}_n \rangle \}$ 
 into the basis $\{ | \Psi_n \rangle \}$ defined by the eigenstates of the full CI as a function of 
 the iteration index $\ell$ for the same system described 
 in Fig \ref{fg:trace}. A larger negative number correspond to in improved small basis. 
 }
 \end{figure}

Figure \ref{fg:determinant} quantifies the convergence of the small basis $\{ | \chi^{S,\ell}_n \rangle \}$ towards the
basis defined by the eigenstates of the CI $\{ | \Psi_n \rangle \}$. For that purpose we define the logarithm of 
the residual subspace projection as
\begin{align}
\label  {eq:determinant}
LR_{sp}=\ln 
\left\{
1-
|Det\left[
 \langle \Psi_n | \chi^{S,\ell}_m \rangle 
\right] 
|^{1/M_S} 
\right\}.
\end{align}
In Eq. (\ref{eq:determinant}) $Det\left[
 \langle \Psi_n | \chi^{S,\ell}_m \rangle 
\right]$ is the Determinant of a square matrix of size $M_S$ formed 
by the overlap $\langle \Psi_n | \chi^{S,\ell}_m \rangle$. The determinant of the matrix is
a complex number of modulus $1$ in the limit when any eigenstate 
$| \Psi_{n<M_S} \rangle$ can be  
written as a linear combination of $| \chi^{S,\ell}_m \rangle$.  Any error in any member of the small basis 
$\{ | \chi^{S,\ell}_m \rangle \}$   reduces the modulus of the determinant by a factor. 
The exponent $1/M_S$ in Eq. (\ref{eq:determinant}) is a standard geometric average.  A large
negative value in Eq. (\ref{eq:determinant}) indicates a very good small basis with a determinant 
that is approaching $1$.

 Figure \ref{fg:determinant} shows the evolution of $LR_{sp}$ given by 
 Eq. (\ref{eq:determinant}) as a function of the iteration 
 index $\ell$ for the same system described in Fig. \ref{fg:trace}. Note that $LR_{sp}$ becomes 
 increasingly negative as a function of $\ell$, which implies a global improvement of the basis 
 approaching  to the one described by the eigenstates of the full CI. 
 
 We next need to characterize how well an individual eigenstate can be described by the small basis. 
 To measure this
 we define the logarithm of the residual projection $LR_n$ as
 \begin{align}
\label  {eq:lrp}
LR_{n}=\ln 
\left[
1-
\sum_{m=0}^{M_S-1}
| \langle \Psi_n | \chi^{S,\ell}_m \rangle |^2 
\right].
\end{align}
Note in Eq. (\ref{eq:lrp}) that, if the normalized eigenstate can be written as a linear combination 
of the small basis $\{  | \chi^{S,\ell}_m \rangle \}$,  the expression in the brackets should be zero. A large
negative number in $LR_{n}$ implies that the eigenstate $| \Psi_n \rangle$  is very well described in 
the small basis.

\begin{figure}
\includegraphics[width=1.00\linewidth,clip=true]{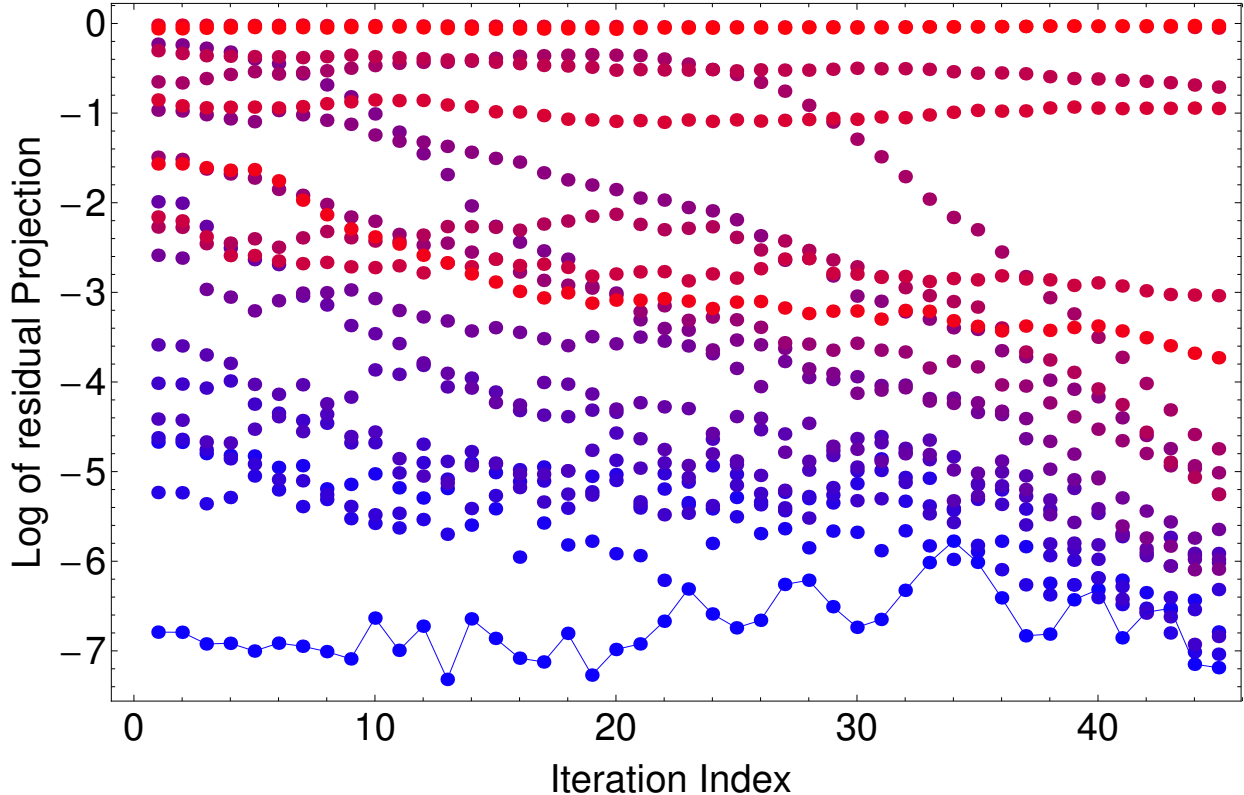}
\caption{
\label{fg:lrp} (Color online) Evolution of the logarithm of residual projection $LR_n$ as a function of the iteration index
$\ell$. Blue (red) color denotes lower (higher) values of $n$. The continuous line follows the ground state eigenstate. 
}
\end{figure}

Figure  \ref{fg:lrp} describes the evolution of $LR_{n}$ for different eigenstates of the CI as a function 
of the iteration index $\ell$. 

 The blue (red) contribution to the color decreases (increases) as the index $n$ increases. The continuous line follows the ground state. One can clearly observe that the
ground state of the CI is already very well described at the initialization stage within the Lanczos-like
setup. As the iteration $\ell$ increases, the small basis describes  the lowest-energy excitations  better while higher excitations 
require more iterations. Note that for 25 iterations 15 eigenvectors are very well described within a basis of 20 states. 
The total cost at this  point is 300 000 individual DMC steps. The calculations of 15 eigenstates with the original SHDMC algorithm
for excited states\cite{rockandroll} would had cost at least twice as much. The current algorithm becomes competitive, in addition,
 if one considers that it is parallel, which allows to distribute this cost in multiple tasks ($M_S$) reducing the time to solution to 2\% as
 compared with the original SHDMC algorithm for excited states. 

Finally, for infinite statistics one could in principle obtain the eigenenergies of $\mathcal{\hat{H}}$  from the eigenvalues of
$u_n$ as $E_n\approx -\ln(u_n)/\beta $. This procedure is known to be inefficient to obtain the eigenenergies which are better described by sampling  $\mathcal{\hat{H}}$ as in the CFDMC approach. 
The off diagonal noise in the 
matrix elements of $\mathcal{\hat{U}}$ has a perverse effect on the magnitude of the eigenvalues. Therefore, while this
method is an efficient one to optimize the basis, it should be combined with other methods to obtain the eigenvalue spectra.


\section{Summary and discussions}
\label{sc:summary}

In this paper we have presented a general framework aimed to calculate thermodynamical properties of many-body 
system in an importance sampling DMC context.~\cite{ceperley80} 

We showed that a many-body basis can be optimized to describe a small subspace  maximizing the  overlap with the
subspace  described by the lowest  eigenstates of the Hamiltonian.  The Helmholtz free energy obtained within this truncated
basis is an upper bound of the exact free energy of a system. The corresponding partition function is a lower bound
of the exact partition function. 

This generalization of the SHDMC method for finite temperature takes advantage of complex 
wave functions that do not have nodal pockets. Accordingly, we avoid the appearance of the artificial  potentials in the standard fixed-node approximation when the amplitude of the importance sampling guiding function is zero. The antisymmetric properties of the wave-function are enforced by
a complex phase. This introduces a complex  contribution to the local energy. The complex local energy is handled in the complex weight
of the walkers. Going beyond the local-time approximation used in Ref. \onlinecite{stonehenge}, the evolution of the complex phase factor for $\beta >0$ is now approximately taken into account in the evaluation local energy. The evolution in $\beta$ becomes exact as the
statistical error is reduced as the sampling increases.

In variance CFDMC,~\cite{ceperley88} the solution remains an 
upper bound of the fermionic free energy. While the CFDMC approach  uses
a bosonic-trial wave function without nodes, SHDMC uses a complex linear combination of anti-symmetric functions without nodal 
pockets. 
The walker distribution is prevented to fall into the bosonic ground state solution by the phase factor of the guiding function, which remains antisymmetric, and introduces an effective potential in the local energy.~\cite{ortiz93,stonehenge} 

The present approach shares many aspects of the SHDMC method for complex wave functions, but it also incorporates a key  advantage of the CFDMC~\cite{ceperley88}
approach: several wave functions are optimized at the same time. In systems where many excitations can be approximated by a single amplitude $\Phi({\bf R})$ and a different phase factor $e^{{\bf i} \phi_{n}({\bf R})}$, a correlated sampling approach that reweighs the walkers in Eq. (\ref{eq:evoltau}) as $W_i\rightarrow W_i \Phi^T_n({\bf R})/\Phi({\bf R})$ and changes the phase contribution to the local energy in Eq. (\ref{eq:ELL}), would save significant time. If that approximation were used, this generalization of the SHDMC method would look very similar in spirit with CFDMC,  the main difference being the use of a complex guiding functions that prevents  the exponential growth of the bosonic ground state.

The displacement of each wave function in the small basis during the DMC process is decomposed into a displacement   within the subspace already 
described by the other elements of the small basis plus a contribution orthogonal to the small subspace. The displacement   included within the basis
is used to improve upon the locality~\cite{locality} and local-time\cite{stonehenge} approximations. The displacement   orthogonal to the small subspace is used
to correct the small basis used in the next iteration.

The  serial orthogonalization step required in the original SHDMC algorithm for excited states~\cite{rockandroll,stonehenge} is avoided with a method that allows the calculation of multiple wave functions in
parallel.  In addition, the complications of inequivalent nodal pockets of excited states~\cite{rockandroll} is avoided using complex trial wave functions without nodes. 

The scaling of the cost of an individual iteration of this method is proportional to $M_S \times N_c$; $M_S$ the size of the small basis
and $N_c$ the number of configurations of the DMC run. The cost of and individual DMC step is  dependent of the number of basis functions $N_b$ and electrons. 

It is well known that as the size of the system increases, the number of basis functions $N_b$ required to maintain a fixed error bar 
for a given eigenstate must increase factorially.  But in practice, the error required for evaluation of thermodynamical properties is determined by $\beta^{-1}$: errors must be much smaller than the temperature of the system. Therefore, as the temperature increases and averages of multiple eigenstates are obtained, 
 the detail required by  calculations of the ground state energies with chemical accuracy is no longer necessary. 
 The present approach can take advantage of the acceleration of the algorithms used to evaluate large numbers of determinants.~\cite{nukala09,clark11} For a very large $N_b$, the cost of these algorithms scales as $N_e \times N_b$.

The total cost is dependent on the physical system and the goal of the calculation. If the goal is to converge the entire  basis or to optimize the Free energy, the bottleneck for convergence is the energy gap $E_{M_S+1}-E_{M_S}$ which determines the convergence of the basis towards the highest eigenstates considered.   Accordingly, in this case, the ideal situation for this method would be a system with ($M_S$) nearly degenerate eigenstates well-separated from the rest of the spectra within an energy scale of $k_b T$.  If the goal, instead, is to converge the small basis so as only the lowest $M_L$ eigenstates are well described, the convergence of the algorithm is much faster
 and it is limited by the number of statistical samples and the exponential decay $e^{-\beta (E_{M_S+1}-E_{M_L})}$.  The cost is reduced as compared with the calculation of eigenstates if one accepts an error in the higher excitations. If one wishes to retain the physics of higher eigenstates in the basis, it is computationally more efficient to 
increase $M_S$ (which increases the cost linearly), instead of improving the basis for the higher excitations which increases the cost exponentially. 

Comparisons of the method with full CI calculations show that  SHDMC can be used to optimize many-body basis
sets to maximize the overlap with the lowest energy excitations of the Hamiltonian. Each eigenenergy obtained 
with this method has  lower quality than those obtained with alternative approaches such as LMMC or the
standard SHDMC for excited states. However, this method could be a useful tool to optimize the basis, minimizing the size of the matrices
used in LMMC and thus reducing the effects of numerical noise in LMMC. 

\acknowledgments
The authors would like to thank J. Krogel and P. R. C. Kent for a critical reading of the manuscript and discussions. 
 This work has been supported by the grant ERKCS92 Materials Science and Engineering division
of Basic Energy Sciences, Department of Energy.   

\appendix
\section{Going beyond the locality and local-time approximations}
\label{ap:beyondloc}

Approximate coefficients for the Slater determinant expansion of the eigenstates of $\mathcal{\hat{H}}$ and $e^{-\beta \mathcal{\hat{H}}}$  
can be obtained from the eigenstates of $\mathcal{\hat{U}}= e^{\hat{J}} e^{- \beta \mathcal{\hat{H}}} e^{-\hat{J}}$ in the $\{ \chi^{S}_n \} $ basis [see Eq. (\ref{eq:JUJ})]. 
But, $\mathcal{\hat{U}}$ is not hermitian,
 since $\mathcal{\hat{U}}^\dagger=e^{2 \hat{J}} \mathcal{\hat{U}}e^{-2 \hat{J}}$. Nevertheless, as long as the Jastrow
 factor operator 
 $e^{-\hat{J}}$ has an inverse, $\mathcal{\hat{U}}$ has a set of right eigenvectors
 $|\chi^{U}_i \rangle = e^{\hat{J}} |\Psi_i \rangle $
 and left eigenvectors $\langle \chi^{U}_i |=  \langle \Psi_i | e^{-\hat{J}}  $. Since $\mathcal{\hat{H}}$ is Hermitian,
 its eigenstates $|\Psi_i \rangle $ are orthogonal, which implies that 
 $\langle \chi^{U}_i |\chi^{U}_j \rangle  = \delta_{i,j} $.
  Within statistical error bars, in the small subspace, the matrix elements of $\mathcal{\hat{U}}$ obtained with
 the basis $|\chi^{S,\ell}_n \rangle $ of the iteration $\ell$ are 
given by Eq. (\ref{eq:JUJ}).  In the first iteration
the matrix elements of $\mathcal{\hat{U}}$ can be obtained directly from the Lanczos-like  
procedure.

 Within the small subspace,   $\mathcal{\hat{U}}$ can be written as
 \begin{align}
 \label  {eq:Uu}
  \mathcal{\hat{U}}=\sum_{i} u_i   |\chi^{U}_i \rangle\langle \chi^{U}_i |.
 \end{align}
 
Since the $u_i$ are also the eigenvalues of $e^{-\beta \mathcal{\hat{H}}}$  their dependence with $\beta$ is exponential.
Thus for an arbitrary $\beta^\prime$ the eigenvalue will be $u_i^{\beta^\prime/\beta}$.

 Provided that the difference 
 $|\Delta\tilde{\chi}^{S,\ell+1}_n \rangle = |\chi^{S,\ell+1}_n \rangle-|\chi^{S,\ell}_n \rangle$ is small [which is always 
 valid for $\beta \rightarrow 0$ see Eq. (\ref{eq:improven})], 
 the dependence in $\beta$ of $ |\chi^{S,\ell+1}_n(\beta) \rangle $ 
 can be approximated as follows: 
Let's first define the operator $\hat{R}=\sum_{n=0}^{M_S-1} |  \chi^{S,\ell}_n \rangle \langle  \chi^{S,\ell+1}_n | $ and its inverse
within the small subspace $\hat{R}^\dagger=\sum_{n=0}^{M_S-1} | | \chi^{S,\ell+1}_n \rangle \langle  \chi^{S,\ell}_n | $.

 Accordingly, the dependence in $\beta^\prime$ of the new basis can be approximated as 
 \begin{align}
 \label  {eq:psibeta}
 \left|\chi^{S,\ell+1}_n \left(\beta^\prime \right) \right\rangle = &  
  \hat{R}^\dagger   \mathcal{\hat{U}}^{{\beta^\prime}/{\beta}} \hat{R} 
  \left|\chi^{S,\ell+1}_n  \right\rangle ,
\end{align}
with  $\mathcal{\hat{U}}^{{\beta^\prime}/{\beta}}$ given by Eq. (\ref{eq:Uu}) replacing $u_i$ by $u_i^{\beta^\prime/\beta}$.
\section{Working with eigenstates of $\mathcal{\hat{U}}$}
\label{ap:eigenstates}
While in some systems eigenstates of $\mathcal{\hat{H}}$ are always real (e.g. confined systems with time reversal symmetry), in many cases
the wave function of the eigenstates is known to be complex, without nodal pockets. In those cases Eq. (\ref{eq:lapltrn}) is valid and no
walker needs to be killed or rejected because the eigenstate wave function does not have a nodal surface.~\cite{stonehenge} In these cases
it might be advantageous to propagate single eigenstates of $  \hat{R}^\dagger   \mathcal{\hat{U}}^{\frac{j}{k}} \hat{R} $, since the variance of the weights is minimized and 
lower (larger) $T$ ($\beta$) can be reached with less statistical data.   An additional advantage of working with functions that are closer to the eigenstates is that the locality and local-time approximations
 can be used. 

Since any $|\chi^{S}_n \rangle = \sum_{i=0}^{M_S-1}  \langle \chi^{X}_i  |\chi^{S}_n \rangle |\chi^{X}_i \rangle$ can be 
written as a linear combination of $|\chi^{X}_i \rangle$ and vise versa,  when eigenstates are complex one can use as trial 
wave function $e^{\hat{J}} |\chi^{U}_i \rangle$ in the SHDMC propagation and sampling. Then the propagation of
 $|\chi^{S}_n \rangle$ can be written as a linear combination of the propagation of the eigenstates of $\mathcal{\hat{U}}$.
as: 
 \begin{align} 
 \label  {eq:conversion}
 |\delta\chi^{S}_n \rangle &= 
 \sum_{i=0}^{M_S}  
 \left\{
 \langle \chi^{X}_i  |\chi^{S}_n \rangle 
 e^{-\beta (E^X_i-E^{S}_n)} \left[
  |\chi^{X}_i \rangle +
 |\delta \chi^{X}_i \rangle 
 \right]
 \right\} -  |\chi^{S}_n \rangle 
 \text{ and } \nonumber \\ 
  |\delta\tilde{\chi}^{S}_n \rangle &= 
 \sum_{i=0}^{M_S}  
 \left\{
 \langle \chi^{X}_i  |\chi^{S}_n \rangle 
 e^{-\beta (E^X_i-E^{S}_n)} \left[
  |\chi^{X}_i \rangle +
 |\delta \tilde{\chi}^X_i \rangle 
 \right]
 \right\} -  |\chi^{S}_n \rangle 
  \end{align}
$E^{S}_n$  results from the condition 
 $\langle \chi^{S}_n  |\delta\chi^{S}_n \rangle = 0$ and it is given by
 \begin{align}
 \label  {eq:ESn}
 e^{-\beta E^{S}_n} = \sum_i 
 \left[
 |\langle \chi^{X}_i  |\chi^{S}_n \rangle|^2+\langle \chi^{X}_i  |\chi^{S}_n \rangle \langle \chi^{S}_n  |\delta\chi^{X}_i \rangle
 \right] e^{-\beta E^X_i} \; ,
 \end{align}
replacing $X$ by $U$ in Eqs. (\ref{eq:conversion}) and (\ref{eq:ESn}). The brackets $\langle \chi^{U}_i  |\chi^{S}_n \rangle$ are obtained by diagonalizing $\mathcal{\hat{U}}$: the coefficient $n$ of the left eigenvectors $i$ in the small 
 basis  $\{ \langle \chi^{S}_n | \}$. 

The disadvantages is that complex eigenstates appear only in certain Hamiltonians or boundary conditions. Albeit without 
nodes, they might have large variation in probability density, in particular for small magnetic fields or twist boundary conditions
close to high symmetry points. Large variations in the probability density hinder correlating sampling. 

\section{Working with eigenstates pairs}
\label{ap:pairs}

It is well known that in many physical systems the energy spacing between eigenstates decreases as the size of the 
system increases. It is also known that the variance of the local energy, which is related to the statistical error in the
energy,  increases as the size of the system increases.~\cite{nemec10} Accordingly as the size of the system increases, it becomes
more difficult to obtain eigenstates. As the size of the system increases the error in the variance introduced by a linear combination of eigenstates
in the Slater part of the wave function becomes smaller than the variance introduced by short range correlations. These
short range correlations cannot be 
accounted by the Slater part, even with a very large basis  $\{| n \rangle \}$, or with simple Jastrow factors. Accordingly, 
in large systems little is lost by using a linear combination of eigenstates, since their contribution to the variance is proportional to the energy separation that decreases as the system become larger. In contrast, much is gained avoiding 
the divergences at the nodes by using complex linear combinations of eigenstates, in particular, to obtain average 
of thermodynamical
properties. 
 However,  to propagate for larger $\beta$ with a branching algorithm, it will be necessary to minimize the variance of the local energy. 

If the 
eigenstates are real, one must use a linear combination of eigenstates. The minimum variance will be reached by 
constructing the Slater part of the guiding wave function with linear combinations of 
eigenstates with consecutive eigenvalues of $u_i$. In this work we use
\begin{align}
\label  {eq:pairs}
|\chi^{V}_{2j-1} \rangle &  = \frac{1}{\sqrt{2}} \left(|\chi^{U}_{2j-1} \rangle+e^{{\bf i} \phi}|\chi^{U}_{2j} \rangle \right) \\
|\chi^{V}_{2j} \rangle &  = \frac{1}{\sqrt{2}} \left(|\chi^{U}_{2j-1} \rangle-e^{{\bf i} \phi}|\chi^{U}_{2j} \rangle \right) \nonumber ,
\end{align}  
where  $e^{{\bf i} \phi}$ is a complex phase  which 
can be adjusted to minimize the variance of the amplitude of the complex wave function $\Phi^T_n({\bf R})$.  
The conjugate vectors  $ \langle \chi^{V}_{2j-1} | $ are constructed using complex conjugate coefficients 
and the left eigenvectors of $\mathcal{\hat{U}}$ in the small basis $\{ \langle \chi^{S}_n | \}$.

Their evolution in imaginary time is given by
\begin{align}
\label  {eq:pairsbeta}
|\chi^{V}_{2j-1} (\beta^\prime) \rangle &  = \frac{1}{\sqrt{2}} \left[(u_{2j-1})^{\frac{\beta^\prime}{\beta}} |\chi^{U}_{2j-1} \rangle+e^{{\bf i}\phi} (u_{2j})^{\frac{\beta^\prime}{\beta}}  |\chi^{U}_{2j} \rangle \right] \\
|\chi^{V}_{2j} (\beta^\prime) \rangle &  = \frac{1}{\sqrt{2}} \left[(u_{2j-1})^{\frac{\beta^\prime}{\beta}} |\chi^{U}_{2j-1} \rangle-e^{{\bf i}\phi} (u_{2j})^{\frac{\beta^\prime}{\beta}}  |\chi^{U}_{2j} \rangle \right] \nonumber
\end{align}


\begin{thebibliography}{99}
\bibitem{fn:back_flow} The theory can be extended to a back-flow operator~\cite{rios06} as long at it has an inverse.
\bibitem{fn:complications} Since we used guiding functions
without nodal pockets, the complications of inequivalent nodal pockets~\cite{rockandroll} are also avoided.
\bibitem{fn:fixedphase} 
The fixed-phase approximation~\cite{ortiz93}, which corresponds to neglecting the term inside 
the $\{ \}$ in Eq. (\ref{eq:EL}), is never used in this work.
\end{thebibliography}
\end{document}